\documentstyle[preprint]{aastex}
\def \CQ             {{(58534)~1997~CQ$_{29}$}}
\def \WW             {{1998~WW$_{31}$}}

\def \QT             {{2001~QT$_{297}$}}

\def \RZ             {{(66652)~1999~RZ$_{253}$}}
\def \ref            {\noindent\hangindent0.5in\hangafter=1}

\begin{document}

\title  {The Orbit and Albedo of Transneptunian Binary \CQ .}

\author{ Keith S.~Noll, Denise C.~Stephens} 
\affil{Space Telescope Science Institute, 3700 San Martin Dr., Baltimore, MD 21218}
\email{noll@stsci.edu, stephens@stsci.edu}

\author{ Will M.~Grundy}
\affil{Lowell Observatory, 1400 W.~Mars Hill Rd., Flagstaff, AZ 86001}
\email{grundy@lowell.edu}

\author{ David J.~Osip}
\affil{Observatories of the Carnegie Institution of Washington, 813
Santa Barbara St., Pasadena, CA 91101}
\email{dosip@lco.cl}

\author{ Ian Griffin}
\affil{Museum of Science and Industry, Liverpool Rd., Castlefield,
Manchester, UK  M3 4FP}
\email{i.griffin@msim.org.uk}

\newpage

\begin{abstract}  


We have measured the separations and position angles of the two
components of the binary transneptunian object \CQ\ at eight epochs.
From these data we are able to constrain the orbit and mass of this
binary system.  The best fitting orbit has an orbital period of $P$ =
312$\pm$3 days. The orbital eccentricity is $e$ = 0.45$\pm$0.03 and the
semimajor axis is $a$ = 8,010$\pm$80 km.  The mass of the system is
surprisingly low at 0.42$\pm$0.02 $\times 10^{18}$ kg.  To account for
the observed brightness consistent with the low mass and a range of
plausible densities, the geometric albedo must be very high; for an
assumed density of 1000 kg m$^{-3}$ we find a red geometric albedo of
$p_R$ = 0.37, an order of magnitude higher than has been generally
assumed for transneptunian objects.  If objects with comparably high
albedos are common in the Kuiper belt, estimates of the mass of the
belt must be significantly reduced.  The semimajor axis of \CQ 's orbit
is 2.8\% of the Hill radius of the primary, a value strikingly similar
to this same ratio in other transneptunian binaries, main-belt
binaries, and regular satellite systems.

\end{abstract}


\newpage
\section{Introduction}

\CQ\ is one of fourteen known transneptunian binaries (TNBs) (Noll
2003).  In addition to \CQ , three others of these fourteen have orbits
that have been determined sufficiently to derive masses, \WW , \QT ,
and \RZ\ (Veillet et al.~2002; Osip et al.~2003; Noll et al.~2004). 
The opportunity to directly measure the mass makes TNBs among the most
attractive objects for study among the $\sim$1000 known transneptunian
objects (TNOs).  

\CQ\ was discovered at the Canada-France-Hawaii Telescope (Chen et
al.~1997) and was found to be a binary object from observations made
with the Hubble Space Telescope in November 2001 (Noll and Stephens
2002).  Follow-up observations made in June and July 2002 showed a
wider separation than at initial discovery and a very small rate of
angular motion.  Our initial conclusions were that an eccentric orbit
was likely (Noll et al.~2002).  With  additional epochs we are now
able to determine an orbit for this binary system yielding a direct
measurement of its mass.   Photometry of the individual components can
be used to constrain the relative masses and radii of the components. 
Albedo as a function of density is also constrained.  In this work we
detail our observations and explore the physical implications of the
mass determined for this binary system.

\section{Observations}

We have been able to constrain the orbit of \CQ\ by combining data
obtained with the Hubble Space Telescope (HST) and with the Magellan
Project Baade and Clay 6.5m telescopes.  The observations are summarized
in Table 1 and described in detail separately below.

\subsection{HST}

HST observations were obtained using the Wide Field Planetary Camera 2
(WFPC2) and Near Infrared Camera Multi-Object Spectrograph (NICMOS) in
three separate observing programs.  All observations employed moving
target tracking with HST to follow the apparent motion of \CQ . 

The observations made with the WFPC2 camera on four separate dates have
been described in detail by Noll et al.~(2002).  The first epoch of
WFPC2 observation was obtained in the WF3 detector with 0.\arcsec 1
pixels; the remaining pointings centered \CQ\ on the higher spatial
resolution, 0.\arcsec 045-per-pixel PC detector.  Three different
filters were used in the WF3 observations, the F555W, F675W, and F814W;
only the F814W was used with the PC.  Two or more undithered exposures
were recorded in each filter to facilitate the removal of cosmic rays.

Three additional observations of \CQ\ have been made using HST's
Near Infrared Camera and Multi-Object Spectrometer (NICMOS) on 8 May
and 16 May 1998 and 23 April 2003.  The May 1998 data were exposed as
target location images for attempted grism spectroscopy of several
bright TNOs (Kern et al.~1999).  On each date three 32 second exposures
of the target using the STEP32 multiaccum sequence were obtained at dithered
positions in the NIC3 camera with the F150W filter in place.  These
relatively short exposures yielded low S/N images of \CQ .  We were
able to reliably detect the secondary only on 16 May 1998, as indicated
in Table 1.  The April 2003 observation of \CQ\ consisted of four
exposures, two each in the F110W and F160W filters (Fig.~1).  Exposures
were 256 sec and 512 sec in the two filters at each of two dithered
positions on the detector.  A STEP256 sample sequence was used with 11
and 12 samples respectively.  HST tracked the apparent motion of the binary
during all NICMOS (and WFPC2) observations and, importantly,
simultaneously corrected for the parallax induced by the telescope's
orbital motion.  These data yielded high S/N detections of both
components of the binary system.  NICMOS data were reduced using the
standard on-the-fly pipeline reduction with the most up-to-date dark
and flat reference files. 


\subsection{Magellan}

Magellan data were obtained with the Raymond and Beverly Sackler
Magellan Instant Camera (MagIC) on eight separate nights in 2002-2004. 
Of these, we have considered three epochs not already sampled by high
quality HST observations, UT 28-29 January 2003, 15 December 2003, and
08-09 March 2004
which offer the best combination of optimal observing conditions and
potential to add unique constraints to the orbit (Fig.~2).  Other dates either
overlapped in time with HST observations or were obtained on nights with
poor seeing.

MagIC is a cryogenic CCD camera with a 2k$\times$2k SITe detector. 
Permanently mounted at a folded port of the Magellan project Clay 6.5m
telescope, MagIC provides a pixel scale of 0.\arcsec 069.  All
observations reported here were obtained with the Sloan $i^{\prime}$
filter in place.  Standard overscan subtraction and flat fielding were
applied to all frames. 

On each date two series of at least three separate 150-300 second
exposures were obtained.  During the integration, the pointing was
fixed (sidereal tracking) so that \CQ\ moved.  In January 2003 during
300 second exposures, the target moved 0.\arcsec 195 at a position
angle of 294.7 degrees.  In December 2003, 300 second exposures
correspond to only 0.\arcsec 07 of sky motion for the target.  Finally,
in March 2004, shorter integrations of 150 seconds resulted in target
image motion of 0.\arcsec 127 at a position angle of 293.1 degrees.
This motion further complicates accurate measurement of the separation
of the two components.

\section{Analysis}

\subsection{HST astrometry}

In order to obtain the most accurate positional information from our
images, we applied a detailed point-spread-function (PSF) fitting
routine to each of the individual images.  Model PSFs were generated
using the TinyTim software package (Krist and Hook 2003).  Following the
method described in Noll et al.~(2002) we iteratively scaled and
shifted synthetic PSF pairs, subtracted from the image, and computed
residuals.  Models with the smallest residuals were used to calculate
the most likely separation and position angle for the two components. 
Errors in centroiding each component are typically of order $\pm$ 0.1
pixel.  The uncertainty in separation and position angle is then
calculated directly from the combined centroiding uncertainties of each
component.  Results are listed in Table 1.  

The positional information derived from our images are the raw data
that we employed in fitting orbits.  In an ideal case where there are
high quality astrometric measurements at several points along an orbit,
including measurements near pericenter and apocenter, orbit analysis
can result in a  precise determination of the system mass.  The
measurements of \CQ\ are, however, far from ideal.  All but one epoch
of our high precision HST data are clustered near position angles of
340 degrees.  The sparseness of the sampling (Table~1, Fig.~2) limits
our ability to  determine the orbit and the mass with the HST data
alone.  The HST data robustly determine the orbital period but do not
fully constrain the remaining orbital elements.  To break this impasse,
we sought additional constraints from ground-based observations.

\subsection{Magellan astrometry}

\CQ\ was observed at Magellan as part of an ongoing program of TNO
observations that resulted in the discovery and characterization of \QT\
(Osip et al.~2003).  Because the ground-based data are considerably
lower in resolution than HST data, we chose to include only two epochs
in our analysis based on their ability to add constructively to our
orbit fitting.

Seeing on the nights of UT 28 January 2003, 15 December 2003, and 09
March 2004 was 0.\arcsec 65, 0.\arcsec 55, and 0.\arcsec 30
respectively.  Because of the small 0.\arcsec 069 pixels, the
point-spread-function (PSF) is oversampled in all images.  While this
leads to a reduction in the signal-to-noise per pixel for any given
source, it is ideal for identifying partially resolved binaries through
comparison of the TNO PSF with stellar PSFs in the same frame.

A 2-D truncated Gaussian is fit to the images of stars to determine
the best-fit FWHM of the PSF for each image.  This PSF is then compared
to the image of the binary.  The images of the binary on the night of
09 March 2004 show two clearly resolved peaks (Fig.~2).  The separation is
measured by fitting PSFs to each peak.  The data from 28 January 2003
do not resolve the binary components, but do indicate elongation of the
stellar PSF corresponding to the target motion vector.  For a
two-component model with a magnitude difference consistent with that
observed at other epochs we can place an upper limit on the separation
at 0.\arcsec 3, or slightly less than half the FWHM of the seeing on
that night.  Data from 15 December 2003 suffer contamination by a
similar magnitude background object within 0.\arcsec 5 - 0.\arcsec 7. 
Nevertheless, adopting a two-component model, these data do 
constrain the separation of the primary and secondary components to
be no greater than 0.\arcsec 3.  

%
%

\subsection{Orbit Fitting}

Orbit fitting was accomplished by iteratively minimizing differences
between observed and model sky-plane offsets between primary and
secondary, projected for the time of each observation.  This procedure
has the advantage of properly accounting for the time variable
projection of the orbit as seen from Earth.  The parallax motion of the
Earth and the seasonally changing aspect of the TNB's orbit both produce
appreciable variations in the projected shape of the orbit.  With our
observations spread over several years, it was essential to include
these effects.  

Two separate solutions to the orbit are found, each comprising equally
good fits to the observed data.  The orbital elements of both
possible solutions are listed in Table 2.  The orbits differ primarily
in the orientation of the orbit plane and in the timing of mutual events
but are otherwise very similar.  We have plotted the solution with the
lower $\chi^2$ in Fig.~3, though we note that the difference is
insignificant. 

Because of the sparseness of the sampling and the clustering of the
highest quality data at a position angle near 340 degrees, the orbit is
dependent on the lowest quality data points we have included in the
fit, the upper limits from 28 January and 15 December 2003.  As
described above, we have attempted to apply conservative criteria to
determining this upper limit in order to insulate the orbit fit from an
error in this single datum.  It is worth noting that even for one-sigma
upper limits to the separation as high as 0.\arcsec 5, the orbit fit
remains essentially unchanged, although with greater uncertainties in
the orbit parameters.  Apparently, even very weak, fractional-sigma
constraints are all that are needed to collapse the uncertainties in
the orbit fits.  It is also possible to consider the extreme case where
we simply eliminate the upper-limit points from our fits.  In that case
the orbit fitting routine again finds essentially the same orbit as
above, but with uncertainties large enough that the eccentricity and
semi-major axis of the orbit become indeterminate, though the orbit
period remains well-determined.  A single orbit fit is replaced by a
family of orbits with fixed $a$ that are able to equally-well fit the
remaining data.  The constrained orbit we find when using all the data
is near the low-mass, low-eccentricity, high probability end of this
family of possible orbits.  Higher masses correspond to increasingly
unlikely orbits having high eccentricity and an orbit plane inclination
approaching an edge-on view.  Further observation, particularly at
times when position angles are away from our cluster of data at 340
degrees, will be able to confirm the validity of our derived orbit.

\subsection{Photometry}

Photometry of the individual components was performed using the same
PSF-fitting technique used to determine relative positions.  The
optimally-scaled PSF is used to determine the flux including
appropriate corrections for finite aperture size.  We have corrected
photometry for phase angle effects using a phase function of 0.15
mags/deg at all wavelengths (Sheppard and Jewitt 2002).  Measured
photometry is listed in Table 3.

Photometric data from five filters can be combined to produce a low
resolution optical-near-infrared spectrum as we have done in Fig.~4. 
As is typical for TNO albedo spectra (e.g. Noll et al.~2000, McBride et
al.~2003) there is a red slope from the optical to approximately 1
micron after which the slope flattens.  The spectral slope in the
optical has been observed to be highly correlated independent of filter
over a range from B through I (430-800 nm; Jewitt and Luu 2001).  The
spectral gradient, $s$, defined as the fractional change in reflectance
per 100 $nm$ (Boehnhardt et al.~2001) is $s=0.19$ and $s=0.14$ for
components A and B respectively measured over the range from F555W to
F814W which approximates V and I.  This falls in the mid-range of V-I
gradients we have measured for 41 non-resonant TNOs (Stephens et
al.~2003).  The gradients we measure for the individual components are
marginally within the error of the V-I gradient for the combined pair,
$s=0.31\pm 0.13$, measured from the ground (Boehnhardt et al.~2001).

Variability is an important issue that must be addressed when combining
photometric data from different epochs.  TNOs are known to have
intrinsic lightcurves, some with significant amplitudes (need a
reference and quantitative number).  As listed in Table 1, the data
obtained with HST at multiple epochs is ill-suited for an evaluation of
possible lightcurve variations.  The F814W filter is the only one used
on multiple dates, a total of four different epochs.  As noted in detail
in Noll et al.~(2002), there is evidence for possible large amplitude
variations in component A (normally the brighter of the pair) with
variations in the I band of as much as a magnitude.  However, with
limited sampling, this observation remains unverified.  Ground-based
observations of the lightcurve of this object could resolve the
question of whether large albedo or shape asymmetries exist on either or
both of the components of this binary.

\subsection{Mass, Density, and Albedo}

Once the orbit is sufficiently well determined that an orbital period
and semimajor axis can be determined, it is possible to retrieve the
total mass of the system using Kepler's third law: $ m_A + m_B = {{4
\pi^2 a^3} / {G P^2}}$.  Both orbit solutions result in a 
system mass of 0.42$\pm 0.02 \times 10^{18} kg$.  This mass is notable
in that it is an order of magnitude lower than the system masses of the
three other binary TNOs that have been measured so far.  

It is possible to use measured photometry to determine the relative
size and mass of the components of \CQ .  The mass ratio is given by
$m_A/m_B = (F_A/F_B)^{3/2}$ conditioned by the assumption that both
components have the same albedo in the relevant photometric band. We
computed this ratio from a weighted average of the flux ratios measured
in the F555W, F675W, and F814W filters with WFPC2 and the F110W and
F160W filters with NICMOS.    Individual masses can be calculated using
this ratio and the derived system mass.  We find $m_A = 0.27\pm 0.03 \times
10^{18} kg$ and $m_B = 0.15\pm 0.02 \times 10^{18} kg$.  Similarly, the
ratio of radii is $r_A/r_B = (F_A/F_B)^{1/2}$.  If we assume a bulk density,
it is possible to derive radii from the masses of the components.  For
$\rho = 1000\ kg\ m^{-3}$ we find $r_A = 40\ km$ and $r_B = 33\
km$.  For different assumed densities the derived radii scale by a
factor of $(\rho /1000\ kg\ m^{-3})^{2/3}$.

A direct consequence of the low system mass is the conclusion that \CQ\
must have a high albedo for any plausible assumed density.  As shown in
Fig.~5, even at an extremely low density of $\rho = 500\ kg\ m^{-3}$
we find a red albedo of $p_R = 0.23$ that is nearly six times higher
than has typically been assumed for TNOs.

\section {Discussion}

\subsection{High Albedos and the Mass of the Kuiper Belt}

As is the case for a single object of measured flux and unknown mass,
the mass of the Kuiper Belt as a whole is proportional to the assumed
albedo to the minus three halves power, independent of assumptions
about the size distribution.  Typically, estimates of the mass of the
belt as a whole have relied on the assumption that $p_R$ = 0.04.  If
the average albedo is five times higher, $p_R$ = 0.2, the mass
calculation overestimates the mass of the belt by more than an order of
magnitude.  For example, Bernstein et al.~(2004) estimate the total
mass of the non-resonant, non-scattered population of the Kuiper belt
to be 0.010 M$_{\oplus}$ using the standard albedo assumption and their
latest estimate of the size distribution.  This already low mass
estimate could be significantly too large if the average albedo of TNOs
is greater than the standard $p_R$=0.04.  Unless the density of \CQ\ is
far lower than any other solar system object known, it indicates that
at least some objects in the Kuiper belt have significantly higher
albedos.  Observations of a significant number of objects at thermal
wavelengths using the Spitzer Space Telescope will constrain this
important physical parameter.

\subsection{The Problem of Orbit Sampling}

The period, $P$, is generally relatively easy to determine, given
observations spanning a substantial fraction, or more, of an orbital
period.  The determination of $a$, however, can be more problematic,
particularly in cases where the orbit is highly eccentric and/or the
normal to the orbit plane is highly inclined.  In these situations the
simple strategy of observing at apocenter and pericenter to directly
determine the quantity $2a$ may be impractical or impossible and, in
any case, requires knowledge of the orbit plane orientation.  For
sparsely sampled orbits, the uncertainty in the orbital eccentricity,
$e$, is also a significant barrier to uniquely determining $a$.  Our
analysis of the orbit of \CQ\ provides a clear example of the
difficulties involved.  Determination of other TNB orbits will
inevitably face similar complications.

\subsection{Eccentricity and the Origin of Binaries}

There is growing evidence that high eccentricity is a common property
of TNBs.  At the present time, four TNBs (excluding Pluto/Charon which
has been circularized due to tidal interactions) have orbits
sufficiently well-determined that their eccentricity is constrained. 
The first TNB to have its orbit measured, \WW , has an eccentricity of
$e$ = 0.82 $\pm$0.05 (Veillet et al.~2002).  Osip et al.~(2003) have
determined that \QT\ has an eccentricity of $e$ = 0.25$\pm$0.1.  We
have recently measured the orbit of \RZ\ and find an eccentricity of
$e$ = 0.46$\pm$0.01 (Noll et al.~2004).  Adding \CQ\ to the mix, with
$e$ = 0.45$\pm$0.03 appears to indicate that moderately high
eccentricity may be the norm for TNBs, though extremely high
eccentricities are not.  This result suggests that the formation of
binaries through the exchange reaction mechanism proposed by Funato et
al.~(2004), which results almost exclusively in eccentricities $e >$
0.8 is not the primary mechanism for the formation of TNBs.  However,
the continuing uncertainty in the formation of stellar binaries
(e.g.~Mathieu 1994), despite more than 200 years of study, is a
sobering reminder of the difficulty of making genetic inferences from
orbital data alone.

\subsection{Transneptunian Binaries and the Hill Radius}

The radius of the Hill sphere, $r_H$, in a three-body system is given
by the equation $ r_H = a (m_A/3M_{\odot})^{1/3}$ where $a$ is the
semimajor axis of the primary's heliocentric orbit.  Orbits within the
Hill sphere are stable relative to the Sun, though perturbations from a
fourth body (e.g. Neptune in the case of TNOs) can alter the zone of
stability.  For binary systems where the mass is known, it is possible
to compute the separation of binary components in terms of the Hill
radius.  The size of the Hill radius is $\sim$7000 times the radius of
the largest component of the binary in the four TNBs  with measured
mass.  The derived orbits of these systems have semimajor axes that are
0.8\% to 6\% of the Hill radius (Table 4).  Pluto has a similarly large
Hill radius, but differs in that Charon orbits at only 0.25\% of this
distance.  TNBs with comparably tight orbits may exist but be
undetected with current searches.  Indeed, the large amplitude
lightcurve of 2001~QG$_{298}$ may be the first recognized contact
binary in the Kuiper Belt and suggests that 10\% to 20\% of the
transneptunian population may consist of similarly close pairs
(Sheppard and Jewitt 2004).

The majority of known binaries in the main asteroid belt orbit much
closer to their primaries in terms of absolute distance and also in
terms of the semimajor axis relative to the size of the primary,
$a/r_A$ (Merline et al. 2003) than known TNBs.  This is
partially due to observational selection effects limiting the discovery of
close TNBs.  However, it is interesting to note that, in terms of the
Hill radius, the semimajor axes of main-belt asteroids are quite
comparable to TNBs, falling in a range of one to a few percent (Table 4).  

We can also compare TNBs to planetary satellite systems.  The four
Galilean satellites orbit in a range from 0.7\% to 3\% of Jupiter's
Hill sphere radius while the most distant satellites, the retrograde
irregulars, orbit at nearly half the Hill radius (Sheppard and Jewitt
2003).  In terms of their separations in units of the Hill radius, TNBs
as well as main-belt binaries are more similar to regular satellites
systems than to the irregular captured objects.  The significance of
this similarity awaits a comprehensive theory of planetesimal binary
formation in the early solar system.

\section{Conclusions}

The orbit of the \CQ\ binary system has been determined with a
combination of data from the Hubble Space Telescope and the Magellan
Observatory.  The derived orbital parameters reveal a system of nearly
equal-sized bodies in a moderately eccentric orbit.  The mass derived
from the orbit is an order of magnitude lower than the three other TNBs
that have been measured to date, and is only $\sim$1/35,000 of the mass
of the Pluto/Charon system.  With probable diameters of less than 80
km, the components of \CQ\ are approaching the size range of the
largest cometary nuclei.  For any reasonable assumed density, these
objects must be remarkably reflective with red albedos greater than
20\%.  If objects this reflective are common in the Kuiper belt,
current mass estimates may be too high by as much as an order of
magnitude.  The separations of transneptunian binaries in terms of the
Hill radius are comparable to the separations of main belt binaries
suggesting possible parallels in their formation.

\acknowledgements {Based on observations made with the NASA/ESA Hubble
Space Telescope. These observations are associated with programs
\#~9060 and \#~9386.  Support for programs \#~9060 and \#~9386 was
provided by NASA through a grant from the Space Telescope Science
Institute, which is operated by the Association of Universities for
Research in Astronomy, Inc., under NASA contract NAS 5-26555.}

\newpage

\noindent{\bf References}

\ref {Bernstein, G.~M., Trilling, D.~E., Allen, R.~L., Brown, M.~E.,
Holman, M., Malhotra, R. 2003 astro-ph/0308467}

\ref {Boehnhardt, H., Tozzi, G.~P., Birkle, K., Hainaut, O., Sekiguchi,
T., Vair, M., Watanabe, J.,  Rupprecht, G., the FORS Instrument Team 
2001,   Astron. \& Astrophys. 378, 653-667}

\ref {Chen, J., Trujillo, C., Luu, J., Jewitt, D., Kavelaars, J.~J.,
Gladman, B., Brown, W.  1997, MPEC 1997-J02}

\ref {Funato, Y., Makino, J., Hut, P., Kokubo, E., 
Kinoshita, D.   2004,  Nature 427, 518-520}

\ref {Kern, S., McCarthy, D., Campins, H., Brown, R.~H., Rieke, M.,
Stolovy, S. 1999, DPS 31.15.03 (abstract)}

\ref {Krist, J.~E., Hook, R.  2003, {\it The TinyTim User's Guide}, v.6.1,
Space Telescope Science Institute, Baltimore}

\ref {Mathieu, R.~D. 1994, Ann.~Rev.~Astron.~Astophys.  32, 465-530}

\ref {McBride, N., Green, S.~F., Davies, J.~K., Tholen, D.~J., Sheppard,
S.~S., Whitely, R.~J., Hillier, J.~K. 2003, Icarus 161, 501-510}

\ref {Merline, W.~J., Weidenschilling, S.~J., Durda, D.~D., Margot,
J-L., Pravec, P., Storrs, A.~D.  2002, in {\it Asteroids}, eds.
W.~F.~Bottke, A.~Cellino, P.~Paolicchi, R.~P.~Binzel, University of
Arizona Press, Tucson, pp. 289-312}

\ref {Noll, K.~S., Luu, J., Gilmore, D.~M., 2000, AJ 119, 970-976.}

\ref {Noll, K., Stephens, D. 2002, IAUC 7824}

\ref {Noll, K., Stephens, D., Grundy, W., Millis, R., Buie, M., Spencer, J.,
Tegler, S., Romanishin, W., \& Cruikshank, D.  2002, AJ 124,
3424-3429}

\ref {Noll, K.~S. 2003, Earth, Moon, \& Planets 92, 395-407}

\ref {Noll, K.~S., Stephens, D.~C., Grundy, W.~M., Griffin, I. 2004,
Icarus, in press}

\ref {Osip, D.~J., Kern, S.~D., Eliot, J.~L. 2003, Earth, Moon, \&
Planets 92, 409-421} 

\ref {Sheppard, S.~S., Jewitt, D.~C. 2002, AJ 124, 1757-1775}

\ref {Sheppard, S.~S., Jewitt, D.~C. 2003, Nature 423, 261-263}

\ref {Sheppard, S.~S., Jewitt, D.~C. 2004, AJ 127, 3023-3033}

\ref {Stephens, D.~C., Noll, K.~S., Grundy, W.~M., Millis, R.~L.,
Spencer, J., Buie, M., Tegler, S.~C., Romanishin, W., Cruikshank,
D.~P.  2003, Earth, Moon, \& Planets 92, 251-260}

\ref {Veillet, C., Parker, J.~W., Griffin, I., Marsden, B.,
Doressoundiram, A., Buie, M., Tholen, D.~J., Connelley, M., Holman,
M.~J. 2002, Nature 416, 711-713}

\newpage

{\tenrm
\null\vskip .1in
\tabskip=1.5em 
\baselineskip=12pt
\tolerance=10000
$$\vbox{ \halign {
# \hfil & #\hfil & #\hfil & #\hfil & #\hfil & #\hfil& #\hfil \cr
\multispan7\hfil{\bf Table 1: Positional Data}\hfil \cr
\noalign { \vskip 12pt \hrule height 1pt \vskip 1pt \hrule height 1pt \vskip 8pt } 
date & instrument/ & separation & pos.~angle & R(AU) & $\Delta$(AU) & phase \cr
     & detector    &  (arcsec)  & (degrees)  &       &              & (deg) \cr
\noalign { \vskip 8pt\hrule height 1pt \vskip 8pt }
\noalign {\bigskip } 
%
%
1998 May 16.11 & NIC3 & 0.36$\pm$0.10 & 355$\pm$15 & 41.243 & 41.047  &1.4 \cr
\noalign {\smallskip}
2001 Nov 17.29 & WF3  & 0.20$\pm$0.03 & 16$\pm$4       &41.554 &41.851 & 1.3 \cr
\noalign {\smallskip}
2002 Jun 18.33 & PC  & 0.337$\pm$0.010 & 334.0$\pm$1.2 &41.608 &41.870 & 1.3 \cr
\noalign {\smallskip}
2002 Jun 30.15 & PC  & 0.334$\pm$0.004 & 337.9$\pm$0.9 &41.611 &42.085 & 1.3 \cr
\noalign {\smallskip}
2002 Jul 12.17 & PC  & 0.331$\pm$0.010 & 340.7$\pm$0.8 &41.614 &42.232 & 1.1 \cr
\noalign {\smallskip}
2003 Jan 28.28 & MagIC & $<$0.3 & -- & 41.665 & 40.869  &  0.8 \cr
\noalign {\smallskip}
2003 May 04.20 & NIC2 & 0.340$\pm$0.005 & 336.6$\pm$1.4 &41.690 &41.186 & 1.2 \cr
\noalign {\smallskip}
2003 Dec 15.5 & MagIC & $<$0.3 & -- & 41.747 & 41.616  & 1.3 \cr
\noalign {\smallskip}
2004 Mar 09.5 & MagIC & 0.38$\pm$0.03 & 338$\pm$5 &41.769 &40.779  &  0.1\cr
\noalign {\vskip 8pt \hrule height 1pt }  
  } }$$}

{
\null\vskip .1in
\tabskip=1.5em 
\baselineskip=12pt
\tolerance=10000
$$\vbox{ \halign {
#\hfil & #\hfil & #\hfil   \cr
\multispan3\hfil{\bf Table 2: Derived Orbital Parameters}\hfil \cr
\noalign { \vskip 12pt \hrule height 1pt \vskip 1pt \hrule height 1pt \vskip 8pt } 
element & orbit solution 1   & orbit solution 2 \cr
        &    & \cr
\noalign { \vskip 8pt\hrule height 1pt \vskip 8pt }
\noalign {\bigskip } 
period (days)        & 312$\pm$3     & 310$\pm$3            \cr
a (km)               & 8,010$\pm$80  & 7,970$\pm$80         \cr
e                    & 0.45$\pm$0.03 & 0.37$\pm$0.01      \cr
i (deg)              & 121.5$\pm$2.0 & 69$\pm$2             \cr
periapse (JD-2450000)& 2,670$\pm$20  & 2,660$\pm$20         \cr
$\theta$ (rad)       & 2.40$\pm$0.07 & 0.31$\pm$0.08        \cr
w (rad)              & 5.41$\pm$0.05 & 5.02$\pm$0.10        \cr
\noalign {\smallskip}
$\chi ^2$            & 2.32          & 2.56                 \cr
\noalign {\smallskip}
mutual events        & $\sim$2136    & $\sim$2041           \cr
\noalign {\smallskip}
system mass (10$^{18}$ kg)  & 0.42$\pm$0.02  & 0.42$\pm$0.02      \cr
\noalign {\bigskip } 
\noalign {\vskip 8pt \hrule height 1pt }  
  } }$$}

{
\null\vskip .1in
\tabskip=1.5em 
\baselineskip=12pt
\tolerance=10000
$$\vbox{ \halign {
\hfil #\hfil & #\hfil & #\hfil & #\hfil & #\hfil & #\hfil  \cr
\multispan5\hfil{\bf Table 3: Photometry}\hfil \cr
\noalign { \vskip 12pt \hrule height 1pt \vskip 1pt \hrule height 1pt \vskip 8pt } 
date &  filter & pivot               & component A & component B  \cr
     &           & $ \lambda (\mu$m) & (mag)       & (mag)        \cr
\noalign { \vskip 8pt\hrule height 1pt \vskip 8pt }
\noalign {\bigskip } 
2001 Nov 17.29 & F555W & 0.55  & 23.85$\pm$0.1   & 24.20$\pm$0.15     \cr
               & F675W & 0.67  & 23.11$\pm$0.15  & 23.52$\pm$0.15     \cr
               & F814W & 0.81  & 22.53$\pm$0.09  & 23.08$\pm$0.15     \cr
2002 Jun 18.33 & F814W & 0.81  & 23.67$\pm$0.03  & 23.23$\pm$0.02     \cr
2002 Jun 30.15 & F814W & 0.81  & 22.75$\pm$0.02  & 23.15$\pm$0.02     \cr
2002 Jul 12.17 & F814W & 0.81  & 22.83$\pm$0.02  & 23.20$\pm$0.02     \cr
2003 May 04.20 & F110W & 1.128 & 22.03$\pm$0.1   & 22.64$\pm$0.1      \cr
               & F160W & 1.606 & 21.50$\pm$0.1   & 22.08$\pm$0.1      \cr
\noalign {\vskip 8pt \hrule height 1pt }  
  } }$$}

{
\null\vskip .1in
\tabskip=1.5em 
\baselineskip=12pt
\tolerance=10000
$$\vbox{ \halign {
#\hfil & #\hfil & #\hfil & #\hfil & #\hfil  \cr
\multispan5\hfil{\bf Table 4: Separations of Binaries}\hfil \cr
\noalign { \vskip 12pt \hrule height 1pt \vskip 1pt \hrule height 1pt \vskip 8pt } 
object &  a (km)    & r$_H$/r$_A$ & a/r$_A$ & a/r$_H$ (\% ) \cr
%
\noalign { \vskip 8pt\hrule height 1pt \vskip 8pt }
\noalign {\bigskip } 
TNBs            &        &      &     &      \cr
\noalign {\smallskip}
\CQ             &  8,010 & 7,050 & 200 & 2.8  \cr
\RZ             &  4,660 & 6,700 &  56 & 0.8  \cr
\QT             & 31,400 & 6,950 & 410 & 5.9  \cr
\WW             & 22,300 & 6,970 & 302 & 4.3  \cr
Pluto           & 19,366 & 6,730 &  17 & 0.25 \cr
\noalign {\bigskip}
Main Belt       &        &      &     &      \cr
\noalign {\smallskip}
(243) Ida       &    108 &  530 &   7 & 1.3  \cr
(45) Eugenia    &  1,190 &  370 &  11 & 2.9  \cr
(762) Pulcova   &    810 &  510 &  12 & 2.3  \cr
(90) Antiope    &    170 &  460 &   4 & 0.9  \cr
(87) Sylvia     &  1,370 &  550 &  11 & 1.9  \cr
(22) Kalliope   &  1,060 &  490 &  11 & 2.3  \cr
\noalign {\bigskip}
Trojan          &  &         \cr
\noalign {\smallskip}
(617) Patroclus &    610 &  790 &  12 & 1.5  \cr
%
\noalign {\bigskip } 
\noalign {\vskip 8pt \hrule height 1pt }  
\multispan5\hfil{Main belt and Trojan data summarized by Merline et al.~2002. }\hfil\cr
  } }$$}

\begin{figure}
\includegraphics[totalheight=0.75\textheight,angle=0]{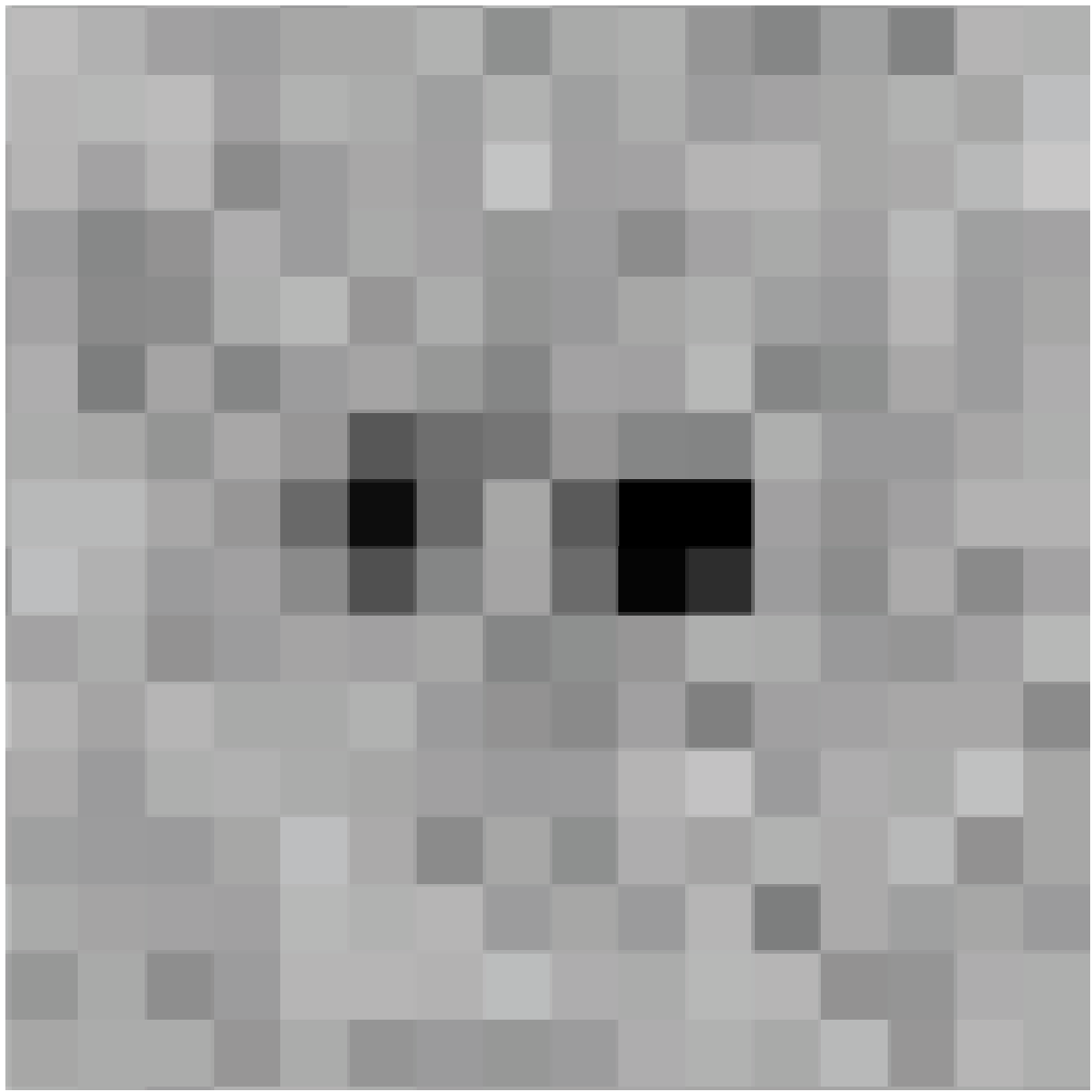}

\caption {A single 512 second exposure with the NIC2 camera and the
F160W filter obtained on 4 May 2003 is shown in the panel on the left. 
The two components of \CQ\ are clearly visible and resolved.  Pixels
are approximately 75 milliarcsec on a side.  }

\label{fig1}
\end{figure}

\begin{figure}
\includegraphics[totalheight=0.75\textheight,angle=0]{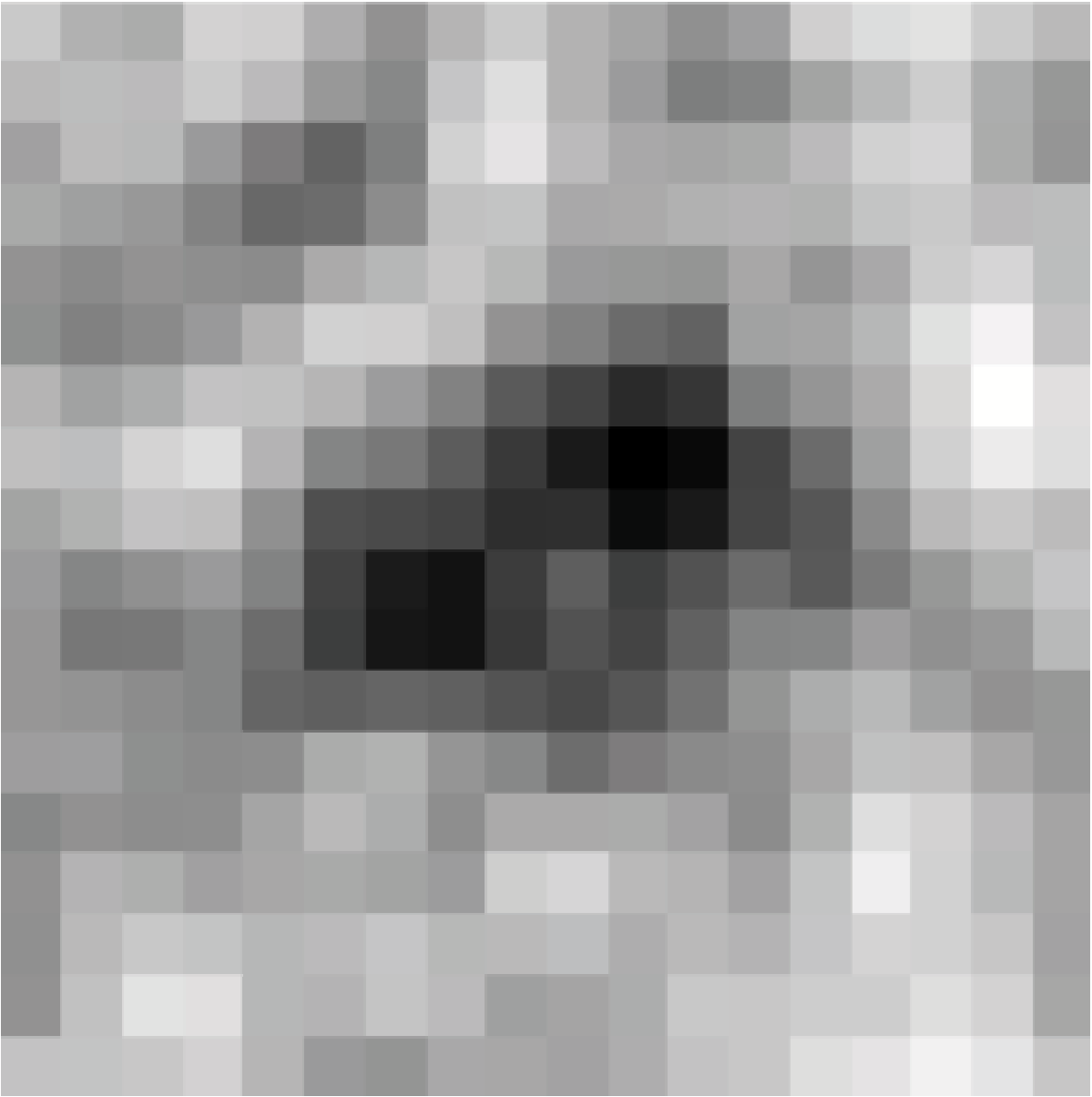}

\caption {Image of \CQ\ taken with MagIC at the Magellan Observatory on
9 March 2004.  Two peaks are clearly resolved in this image.  Pixels are
69 milliarcsec on a side.  This image was created by smoothing the raw
image with a gaussian with a FWHM $\sim$ 2.5 pixels to minimize
background noise. }

\label{fig2}
\end{figure}

\begin{figure}
\includegraphics[totalheight=0.75\textheight,angle=0]{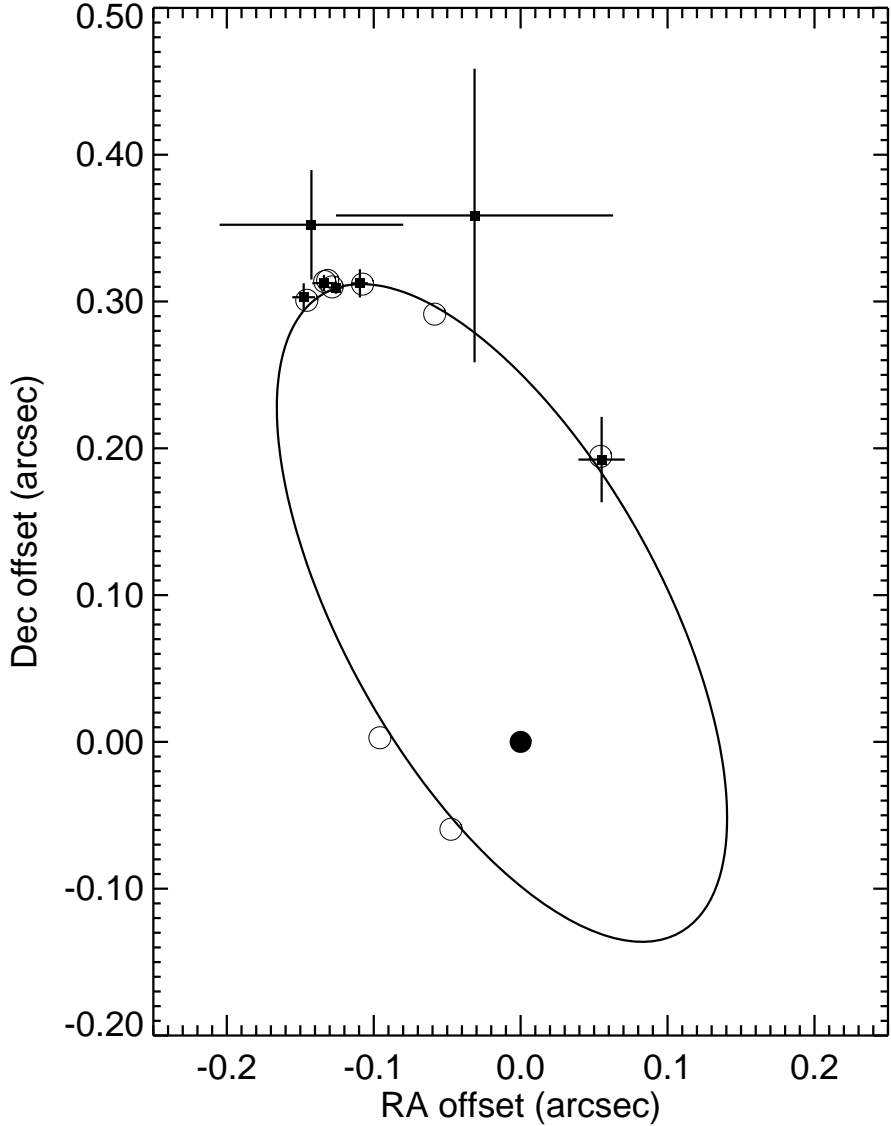}

\caption {The measured separations of component B relative to component A of the
\CQ\ binary are shown as solid points with the combined astrometric uncertainty
indicated at the component B position.  The ellipse is the instantaneous
projection of the best-fitting orbit at an arbitrary time near the
mid-point of our observations.  The actual orbital trace is complex due
to parallax and the precession of the orbit plane.  The expected
position of component B at each epoch is denoted by an open circle. 
Because of the aforementioned changes in geometry, these do not all lie
on the plotted ellipse.  The two points without corresponding
measurements are the predicted separations at the times we recorded upper
limits from MagIC.  All other points are detections.  }
\label{fig3}
\end{figure}

\begin{figure}
\includegraphics[totalheight=0.65\textheight,angle=0]{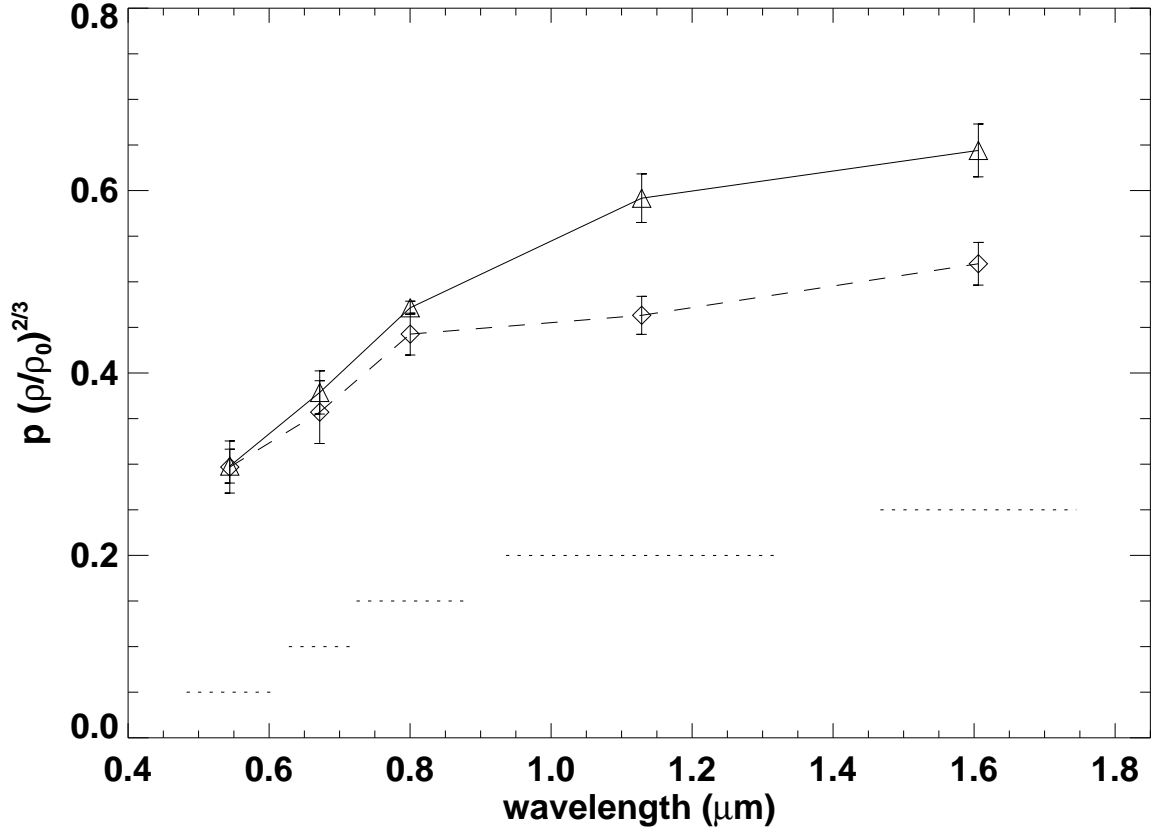}

\caption {The albedo of each component of \CQ\ is plotted at each of
five wavelengths measured.  The curves shown assume that both
components share a bulk density of $\rho_0$ = 1000 kg m$^{-3}$. The
albedo scales with density to the 2/3 power.  The solid curve
corresponds to the larger member of the binary, component A, and the
dotted curve to the smaller component B.  The dotted bars at the bottom
of the figure are show the bandpasses of the five filters used. 
The discrepant F814W photometry from 18 June 2002 was not included in
computing the average flux at this wavelength.
\label{fig4}} \end{figure}

\begin{figure}
\includegraphics[totalheight=0.6\textheight,angle=0]{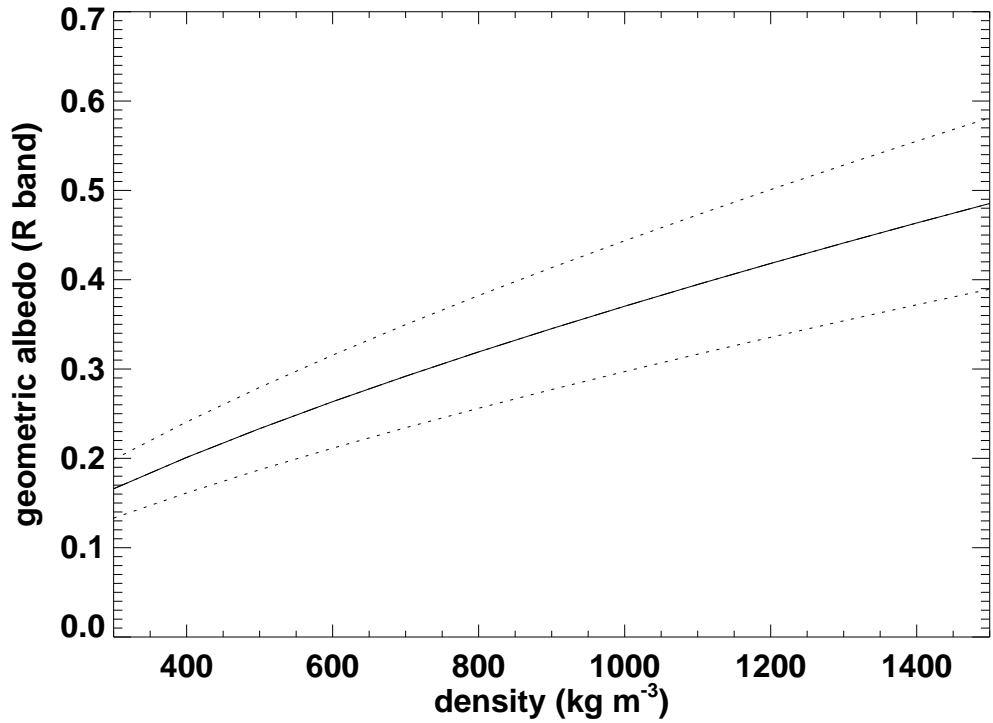}

\caption {Plot of albedo at 675 nm as a function of assumed density. 
The solid curve is computed using the average flux measured at 675 nm
with the uncertainties bounded by the dotted curves. 
For a density  $\rho$ = 1000 kg m$^{-3}$, the average albedo is
$p_{675}$ = 0.39, far higher than generally assumed for transneptunian
objects.  If albedos as high as this are typical, the mass of the
Kuiper Belt may be currently overestimated by an order of magnitude. 
\label{fig5}} \end{figure}

\end{document}